\documentclass[useAMS,usenatbib]{mn2e}
\usepackage{graphicx}
\def\deg{{\rm o}}

\title{Halo-model Analysis of the Clustering of Photometrically Selected Galaxies from SDSS}

\author[Ross \& Brunner]{Ashley J. Ross$^1$ \& Robert J. Brunner$^{1,2}$\\
$^1$Department of Astronomy, University of Illinois, 1002 W Green St., Urbana, IL 61801\\
$^2$National Center for Supercomputing Applications, Champaign, IL 61820}

\date{Accepted to MNRAS}

\pagerange{\pageref{firstpage}--\pageref{lastpage}} \pubyear{2009}

\begin{document}

\label{firstpage}

\maketitle



\begin{abstract}
We measure the angular 2-point correlation functions of galaxies, $\omega(\theta)$, in a volume limited, photometrically selected galaxy sample from the fifth data release of the Sloan Digital Sky Survey.  We split the sample both by luminosity and galaxy type and use a halo-model analysis to find halo-occupation distributions that can simultaneously model the clustering of all, early-, and late-type galaxies in a given sample.  Our results for the full galaxy sample are generally consistent with previous results using the SDSS spectroscopic sample, taking the differences between the median redshifts of the photometric and spectroscopic samples into account.  We find that our early- and late-type measurements cannot be fit by a model that allows early- and late-type galaxies to be well-mixed within halos.  Instead, we introduce a new model that segregates early- and late-type galaxies into separate halos to the maximum allowed extent.  We determine that, in all cases, it provides a good fit to our data and thus provides a new statistical description of the manner in which early- and late-type galaxies occupy halos.  
\end{abstract}


\section{Introduction}
The `halo-model' (see, e.g. \citealt{Kauf97,PS00,CooSh02,zheng05,Tinker}) has been developed to allow one to precisely model the clustering of galaxies.  One can fill dark matter halos with galaxies based on a statistical `halo-occupation-distribution' (HOD), allowing one to model the clustering of galaxies within halos (and thus non-linear scales) while providing a self consistent determination of the bias at linear scales.  Thus, as shown by, e.g., \cite{Z04}, \cite{Blake}, \cite{Tink08}, one can use measurements of galaxy 2-point correlation functions to constrain the HOD of different sets of galaxies and gain information on the nature in which galaxies occupy dark matter halos.  

Such a halo-model analysis can be particularly useful in constraining the clustering of early- and late-type galaxies.  It has long been known that early-type galaxies cluster more strongly than late-type galaxies (recent studies include, e.g., \citealt{W98,N02,Ma03,Z05,R06,Cr06}), and that there exists a corresponding morphology-density relationship \citep{Dre80} which essentially states that the fraction of early-type galaxies increases with the density of the local environment.  Zehavi et al. (2005; hereafter Z05) have incorporated this relationship into their halo modeling (by allowing the fraction of late-type galaxies to decrease as a function of halo mass) and shown that this approach can indeed effectively model the clustering of early- and late-type galaxies.

Recent studies have determined that the morphology-density relationship can be more accurately described as a color-density relationship.  \cite{Ball08} find no residual relation between density and morphology when removing color (but do find a strong residual in density and color when removing morphology) and \cite{Skibba08} find a strong environmental dependance on color, even for fixed morphology.  This implies that deep photometric surveys (which are likely to have little morphological information) should be sufficient for quantifying the clustering as a function of galaxy type.

In this work, we use galaxies that are photometrically selected from the Sloan Digital Sky Survey (SDSS) fifth data release (DR5) to constrain HODs.  \cite{Blake} have previously used photometric data to constrain the HODs of luminous red galaxies (LRGs), and we follow a similar approach to constrain the HODs of early- and late-type galaxies.  The wealth of quality photometric data allows us to precisely constrain the HODs of early- and late-type galaxies at higher redshifts than previous SDSS studies (e.g. Z05). 

Our paper is outlined as follows:  \S\ref{sec:HOD} describes how we use the halo model to obtain model angular 2-point correlation functions of galaxies; \S\ref{sec:data} describes how we both select galaxies from the SDSS DR5 photometric catalog to produce a volume limited sample (to $z = 0.3$), which we further subdivide by type and luminosity, and also estimate the redshift distribution of these galaxy samples; \S\ref{sec:meas} describes how we measure the 2-point correlation functions of galaxies and how we estimate the error on these measurements; \S\ref{sec:res} presents the results of our 2-point correlation function measurements and the best-fit HOD for each galaxy sample; in \S\ref{sec:dis} we compare our results to previous studies and discuss the implications of our measurements; finally, we conclude in \S\ref{sec:con}.  Throughout this work, we assume a flat cosmology with $\Omega_m = 0.28$, $h = 0.7$, $\sigma_8 = 0.8$, $\Gamma = 0.15$ (as used in Ross et al. 2007; hereafter R07).

\section{Halo Modeling}
\label{sec:HOD}
We use the halo model to produce model angular 2-point correlation and cross-correlation functions.  The most basic component of the model is the number density of halos at redshift $z$ with mass $M$, $n(M,z)$.  We determine both $n(M,z)$ and the bias of these halos, $B(M,z)$, by using an ellipsoidal collapse model (e.g., \citealt{Sheth01}) coupled with the methods described in \cite{Nish06}.   We model the probability distribution of the number of galaxies occupying a halo of a given mass, the {\it halo-occupation-distribution} (HOD), to determine the mean number of galaxies, $N(M)$, occupying halos of mass $M$.  Following \cite{zheng05} and \cite{Blake}, we assume separate mean occupations for central galaxies, $N_{c}(M)$ and for satellite galaxies, $N_{s}(M)$.  Thus,
\begin{equation}
N(M) = N_c(M)\times(1 + N_s(M))
\end{equation}
Note what this implies: for a given halo mass, the mean number of satellites is $N_c(M)N_s(M)$, but these satellites are broken between halos which have a central galaxy and those that do not.  The halos that do have a central galaxy have $N_s(M)$ satellite galaxies, and those without a central galaxy have zero satellites.  This reasoning is crucial to understanding the equations presented throughout this section.

Coupling this model with a formalism describing how galaxies distribute themselves within halos allows us to model the power spectrum (which we Fourier transform and then convert to $\omega(\theta)$ via Limber's equation; \citealt{lim}).  We assume that the spatial distribution of satellite galaxies follow the \cite{NFW} dark matter halo density profile:  
\begin{equation}
\begin{array}{lll}
 \rho(M,r) =  & \frac{M}{(cr/r_{vir}(M))(1+cr/r_{vir}(M))^2} & ~\\ \\
 ~ &\times  \frac{1}{4\pi(r_{vir}(M)/c)^3\left[{\rm ln}(1+c)- c/(1+c)\right]} & ~
\end{array}
\end{equation}
where $r_{vir}$ is the virial radius and $c$ is the concentration parameter.  Following \cite{Z04}, we define 200 as the critical over-density for virialization and can thus express the virial radius, $r_{vir}$ as a function of mass as
\begin{equation}
r_{vir} = \left( \frac{3M}{200\times4\pi  \bar{\rho}}\right)^{\frac{1}{3}}
\label{eq:mvir}
\end{equation}
and inversely the virial mass as a function of scale as
\begin{equation}
M_{vir}(r) = 200 \times \frac{4}{3}\pi r^3 \bar{\rho}
\label{eq:mvir}
\end{equation}
where $\bar{\rho}$ is the mean co-moving background density of the Universe.  Using this definition for the virial mass, \cite{Z04} determined via \cite{Bull01} that $c$ can be expressed
\begin{equation}
c(M,z) = 11/(1+z) (M/M_c)^{-0.13}
\end{equation}
where we determine ${\rm log_{10}} (M_c) = 12.49$, where $M_c$ is in units $(h^{-1} M_{\odot})$, for our assumed cosmology (see, e.g. \citealt{Blake}).  We use the Fourier transform of the full (un-truncated) $\rho(M,r)$, $u(k|M)$,  to calculate the power spectrum, and we use the form presented in \cite{scoc01}.  We note that the difference in using the \cite{scoc01} parameterization, and one in which
$\rho(M,r)$ is truncated (using, e.g., Equation 8 from \citealt{jain03} with $r_{vir}$ as upper bounds on
the integrals) is negligible ($< 0.1\%$) in our model angular correlation functions.

\subsection{Modeling the Power-Spectrum}
The equations presented above allow us to model the power spectrum as having a contribution due to galaxies in two separate halos (the 2-halo term) and a contribution due to galaxies that reside in a single halo (the 1-halo term). 
\begin{equation}
P(k,r) =  P_{1h}(k) + P_{2h}(k,r)
\end{equation}
where $P_{1h}(k)$ is split into two components --- one being the power-spectrum due to central-satellite pairs, $P_{cs}(k)$, and the other due to satellite-satellite pairs, $P_{ss}(k)$: 
\begin{equation}
P_{cs}(k) = \int^{\infty}_{M_{vir}(r)} dM n(M) N_c(M)  \frac{2  N_s(M ) u(k|M)}{n^2_{g}}
\label{eq:cs}
\end{equation}

\begin{equation}
P_{ss}(k) =\int^{\infty}_{0} dM n(M) N_c(M) \frac{ \left(N_s(M)u(k|M)\right)^2}{n^2_{g}}
\label{eq:ss}
\end{equation}

\noindent and
\begin{equation}
\begin{array}{ll}
P_{2h}(k,r) & =  P_{matter}(k) \\
~ & \times \left [ \int_0^{M_{lim}(r)} dM n(M) b(M,r) \frac{N(M)}{n_g^{\prime}}u(k|M) \right]^2 
\end{array}
\end{equation}
where $P_{matter}$ is the matter power-spectrum determined via the fitting formulae of \cite{Smith}.  The parameter $M_{lim}(r)$ is the mass limit due to halo-exclusion, which we determine using the methods described by \cite{Tinker} and \cite{Blake}.  The average number density of galaxies is given by $n_{g}$, and $n^{\prime}_{g}$ is the {\it restricted} number density of galaxies.  The two number densities can be expressed as
\begin{equation}
n_{g} = \int_0^{\infty}{\rm d}Mn(M)N(M)
\label{eq:ng}
\end{equation}
and
\begin{equation}
n^{\prime}_{g} = \int_0^{M_{lim}(r)}{\rm d}Mn(M)N(M)
\end{equation}
The scale dependent bias, $b(M,r)$, can be expressed \citep{Tinker} as a function of the halo bias as
\begin{equation}
b^2(M,r) = B^2(M) \frac{[1+1.17\xi_m(r)]^{1.49}}{[1+0.69\xi_m(r)]^{2.09}}
\end{equation}
 where $\xi_m(r)$ is the non-linear real-space matter 2-point correlation function, determined by Fourier transforming the matter power spectrum.  
 
In many cases, it will be useful to calculate the effective mass, $M_{eff}$, given by:
\begin{equation}
M_{eff} = \frac{1}{n_g} \int {\rm d}M M n(M)N(M)
\end{equation}
and also the overall bias of the galaxies given by the model, $b_{gal}$ given by:
\begin{equation}
b_{gal} = \frac{1}{n_g} \int {\rm d}M B(M) n(M)N(M) 
\end{equation}
These parameters are quite useful when comparing our results to each other and also to previous studies.
 
\subsection{Halo Occupation Distribution Model}
We model the HOD as a power-law with a softened transition for both the central and satellite galaxies (note that this implies the softening effect is squared for the satellite galaxies).  This is expressed as
\begin{equation}
N_c(M) = 0.5 \left[ 1 + {\rm erf}\left(\frac{{\rm log_{10}} (M/M_{cut})}{\sigma_{cut}}\right)\right]  
\label{eq:HOD}
\end{equation}
\begin{equation}
N_s(M) = 0.5 \left[ 1 + {\rm erf}\left(\frac{{\rm log_{10}} (M/M_{cut})}{\sigma_{cut}}\right)\right] \times  \left(\frac{M}{M_0}\right)^{\alpha}  
\label{eq:HOD}
\end{equation}
These Equations are entered into Equation 1 to determine the mean occupation of halos at a given mass.

The HOD model has four free parameters, but one can be removed by requiring that $n_g$ match the observed number density of galaxies.  To ensure this, we determine $M_{cut}$ for any chosen combination of $\sigma_{cut}$, $M_0$, and $\alpha$.  We also model the number of early- and late-type galaxies by expressing the fraction of late-type centrals and satellites as a function of halo mass (similar to Z05).
\begin{equation}
f_c(M) = f_{c0} ~ {\rm exp}\left[\frac{ -{\rm log_{10}}(M/M_{cut})}{\sigma_{cen}}\right]
\label{eq:cent}
\end{equation}
and
\begin{equation}
f_s(M) = f_{s0} ~ {\rm exp}\left[\frac{ -{\rm log_{10}}(M/M_{0})}{\sigma_{sat}}\right]
\label{eq:sat}
\end{equation}
where we cap $f_s(M)$ and $f_c(M)$ (which we will from here on express as $f_s$ and $f_c$ in order to be concise) such that they are never greater than 1.

This model once again has four free parameters, but we can remove one since we know the overall fraction of late-type galaxies.  We thus calculate the required $f_{c0}$ for every allowed combination of $f_{s0}$, $\sigma_{cen}$, and $\sigma_{sat}$.

In a previous work (Z05), color selected ``red" and ``blue" galaxies were assumed to be well mixed within halos.  We find that this type of model does not provide an adequate fit to our measurements of the auto-correlations of  early- and late-type galaxies.  We instead assume that if a central galaxy is a certain type, its satellite galaxies will be the same type, up to the extent allowed by the $N(M)$ and $f(M)$ statistics of the given model.  In order to be concise, we express the $P(k)$ in terms of a new function $\Theta(k,M)$:  
\begin{equation}
\begin{array}{ll}
P_{type}(k) = & ~ \\
\int^{\infty}_{M_{min}(r)} dM \Theta(k,M) n(M) N_c(M) N_s(M ) u(k|M)/n^2_{g,type}, & ~ \\
\end{array}
\end{equation}
where $type$ can be either early or late.  Equations \ref{eq:cs} and \ref{eq:ss} now become dependent on type and the relative values of $f_c$ and $f_s$ (which are themselves dependent on mass).  If the late-type satellite fraction is greater than the late-type central fraction, all satellite galaxies around late-type central galaxies will be late-types (and thus each late-type central will have $N_s$ late-type satellites).  If the opposite is true, each early-type central galaxy only has early-type satellites.  For the average late-type central, the fraction of its satellites that are also late-type is the total fraction of late-type satellite galaxies ($f_s$) divided by the fraction of late-type central galaxies  ($f_c$).  Thus, the central-satellite term for late-types can be expressed as:
\begin{equation}
\begin{array}{lll}
\Theta_{cs, late}(k,M) &=  2 f_c & ( f_s > f_c) \\
~& = 2 f_c \times f_s/f_c = 2 f_s & ( f_s < f_c)
\end{array}
\label{eq:cslseg}
\end{equation}

For the central-satellite term for early-type galaxies, we must take into account the fact that there will be late-type satellite galaxies around early-type centrals if $f_s > f_c$.  In this case, we need to determine the fraction of early-type satellites around early-type centrals.  The total fraction of early-type satellites is just $1 - f_s$.  Thus the average fraction of satellites around early-type centrals that are early-type is  $(1 - f_s)/(1-f_c)$ and the total contribution due to early-type satellites around early-type centrals is $(1-f_c)(1 - f_s)/(1-f_c)$.  For $f_c > f_s$, all satellites around early-type galaxies are early-type and each early-type central thus has $N_s$ early-type satellites.  The central-satellite term for early-types can therefore be expressed as:
\begin{equation}
\begin{array}{lll}
\Theta_{cs, early}(k,M)&  = 2 (1 - f_s)  & (f_s > f_c) \\
~ & =2(1 - f_c)  & (f_s < f_c)
\end{array}
\end{equation}

For the satellite-satellite terms, the same logic applies.  If $f_s > f_c$ there will be a term for both late-type satellites around late-type centrals and late-type satellites around early-type centrals.  Around late-type centrals, all satellites are late-type, and the fraction of late-type satellites is just 1.  The total fraction of late-type satellites around early-type satellites is $f_s - f_c$ and the fraction around only halos with early-type central galaxies is thus $(f_s - f_c)/(1-f_c)$.  This term must be squared to account for the total number of late-type satellite-satellite pairs around early-type galaxies.  The total contribution due to halos with early-type centrals is thus $(1-f_c)\times\left[(f_s - f_c)/(1-f_c)\right]^2$.  If instead $f_c > f_s$, there are only late-type satellite galaxies around late-type centrals, thus the contribution is $f_c\times(f_s/f_c)^2$.  We therefore express $\Theta_{ss,late}$ as:
\begin{equation}
\begin{array}{ll}
\Theta_{ss, late}(k,M) =& ~ \\
  \left[f_c+ (f_s - f_c)^2/(1 - f_c) \right] N_s(M ) u(k|M) & (f_s > f_c) \\
  (f_s^2/ f_c) N_s(M ) u(k|M) & ( f_s < f_c)
\end{array}
\end{equation}

For the early-type satellite-satellite term, if $f_s > f_c$, there are only early-type galaxies around early-type centrals.  The fraction of satellites around early-type centrals that are early-type is $(1 - f_s)/(1-f_c)$ and thus the contribution to the satellite-satellite term is $(1-f_c)\times[(1 - f_s)/(1-f_c)]^2$.  For $f_c > f_s$, the fraction of satellites around early-type centrals that are early-type is just 1 and the fraction around late-type centrals is $(f_c - f_s)/f_c = (1 - f_s/f_c)$.  We thus express $\Theta_{ss,early}$ as:
\begin{equation}
\begin{array}{ll}
\Theta_{ss, early}(k,M) = ~\\
   \left[(1 - f_s)^2/(1 - f_c)\right] N_s(M ) u(k|M) &  (f_s > f_c) \\
   \left[1-f_c + f_c\left(1 - f_s/f_c\right)^2\right]N_s(M ) u(k|M) & (f_s < f_c)
\end{array}
\label{eq:sseseg}
\end{equation}

The requirements of Equations 20 through 23 segregate the early- and late-type galaxies as much as possible while maintaining the statistics of Equations of \ref{eq:cent} and \ref{eq:sat}.  This essentially results in a model where smaller mass halos will only be occupied by early- or late-type galaxies and larger mass halos will have a central early-type galaxy, many early-type satellite galaxies and room for a smaller (but significant) number of late-type galaxies.  

In the case of a cross-correlation, one must substitute $2n_{early}n_{late}$ for $n^2_{g,type}$.  For both the central-satellite and satellite-satellite terms, only halos with early-type centrals will contribute if $f_s > f_c$ and only halos with late-type centrals will contribute if $f_c > f_s$.  Thus,
\begin{equation}
\begin{array}{lll}
\Theta_{cs, {\rm el}}(k,M) &=    2 (f_s - f_c) &( f_s > f_c) \\
~ & = 2 (f_c - f_s) &  (f_s < f_c)
\end{array}
\end{equation}

\begin{equation}
\begin{array}{ll}
\Theta_{ss,  {\rm el}}(k,M) = & ~ \\
  2 \left[(1-f_s) (f_s - f_c)/(1 - f_c)\right]  N_s(M ) u(k|M) & (f_s > f_c) \\
2 f_s(1 - f_s/f_c) N_s(M ) u(k|M) & (f_s < f_c).
\end{array}
\end{equation}
These sets of equations account for all of the pairs of galaxies that were present when the auto-correlation of the full sample was measured.  Thus, the expressions that represent the fractions of pairs contributing to each term must add to one (i.e., $\Theta_{cs,late}/2 + \Theta_{cs,early}/2 + \Theta_{cs,el}/2 = 1$ and $\Theta_{ss,late}/(N_s(M ) u(k|M)) + \Theta_{ss,early}/(N_s(M ) u(k|M)) + \Theta_{ss,el}/(N_s(M ) u(k|M)) = 1$).  Inspection of our $\Theta$ expressions reveals that this is indeed the case.

\subsection{Transformation to Angular Correlation Function}
In order to compare our measurements to our HOD model, we must Fourier transform the model power spectra to a real-space correlations function:
\begin{equation}
\xi(r) = \frac{1}{2\pi^2}\int_{0}^{\infty}{\rm d}k~ P(k,r)k^2 \frac{{\rm sin}~ kr}{kr}
\end{equation}
and use Limber's equation to project the real-space model to angular space (assuming a flat Universe):
\begin{equation}
\begin{array}{ll}
\omega(\theta) = &~\\ 
 2/c \int_{0}^{\infty}{\rm d}z ~H(z)(dn/dz)^2 \int_{0}^{\infty}{\rm d}u~\xi(r= \sqrt{u^2+x^2(z)\theta^2}) & ~
\end{array}
\label{eq:Lim}
\end{equation}
where $c$ is the speed of light, $H(z)$ is the expansion rate of the Universe, $dn/dz$ is the normalized redshift distribution, and $x(z)$ is the comoving distance to redshift $z$.

\section{Data}
\label{sec:data}
The data analyzed herein were taken from the SDSS DR5 \cite{Ab05}.  This survey obtains wide-field CCD photometry \cite{C} in five passbands ($u,g,r,i,z$; e.g., \citealt{F}).  The entire DR5 represents close to 8,000 square degrees of observing area.  We selected galaxies lying in the Northern, contiguous portion of the SDSS from the DR5 {\tt PhotoPrimary} database and matched them to galaxies from the DR5 {\tt PhotoZ} table.  We constrained the sample (using the \citealt{Sc} dust maps) to have reddening-corrected magnitudes in the range $18 \leq r <  21$.  We further masked our data by using the same pixelized mask of R07 (which cut on the SDSS DR5 survey area, seeing $>$ 1.$^{\prime\prime}$5, $r$-band reddening $>$ 0.2, bad pixels, satellite trails, etc.).  This left 5,407 deg$^2$ of observed sky.  Following the methods outlined in \cite{Bud03}, we created a volume limited sample with $z < 0.3$ and $M_r < -19.5$ (we note that this same volume limited sample is used in R07).  After masking, this sample, hereafter denoted as $Z3$, contains nearly four million objects (3,980,652).

We also subdivide these data samples by luminosity and type.  We select galaxies with $M_r < -20.5$ from $Z3$ to produce a sample with just over 1.3 million galaxies (1,302750), which we denote $Z3B$.  Each sample ($Z3$ and $Z3B$) is also split by galaxy type based on their type values from the DR5 {\tt PhotoZ} table.  As in R07, galaxies with $t >$ 0.3 are put in our late-type sample and those with $t \leq 0.3$ are put in our early-type sample.  The $Z3$ has nearly as many late-type galaxies (1,984,021) as early-type galaxies (1,996,631), while $Z3B$ has significantly more early-type galaxies (820,789 to 481,961).  In total, this gives six galaxy samples for which we measure $\omega(\theta)$ and determine a best-fit halo occupation model. 

\subsection{Redshift Distributions}
\label{sec:dndz}
\begin{figure}
\includegraphics[width=84mm]{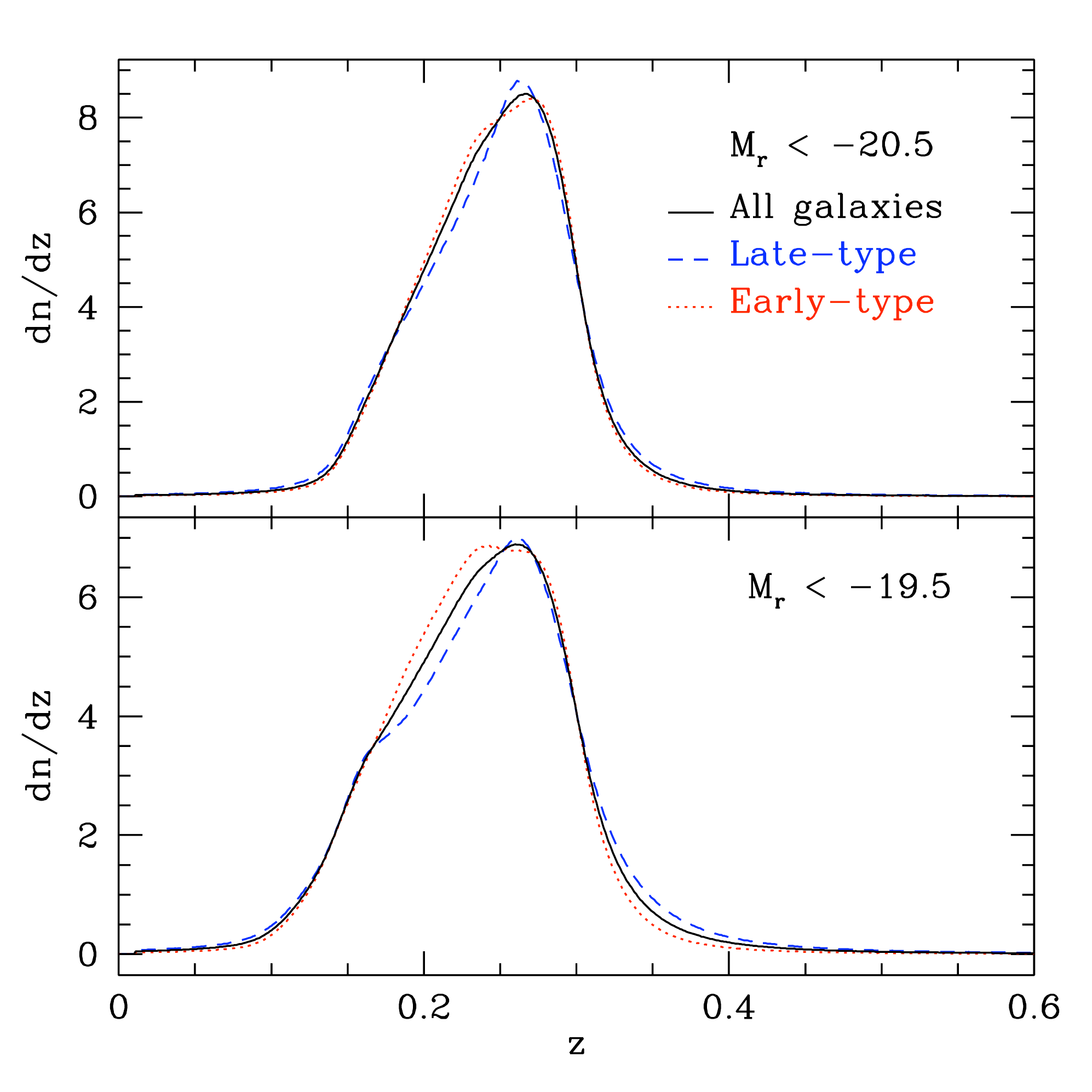}
\caption{The normalized redshift distributions for each of the six galaxy samples we use ($z < $0.3, $M_r < -19.5$, bottom; $z < 0.3$, $M_r <$ -20.5, top) with the distribution for each full sample plotted in black, the late-type distributions plotted in blue, and the early-type distributions plotted in red.}
\label{fig:dndz}
\end{figure}

We require a knowledge of the redshift distribution for each of our galaxy samples in order to compare our observations to theoretical models.  To build the redshift distribution for each sample, we treat each observed redshift as a Gaussian probability-density-function (PDF) with $\sigma$ equal to the estimated error.  The PDFs for each redshift were sampled in order to find the expected number of objects within bins of width 0.001 in $z$.  We normalize these distributions to have unit area and use this result in Equation \ref{eq:Lim}.  The normalized ${\rm d}n/{\rm d}z$ of our samples are plotted in Figure \ref{fig:dndz}.  The normalized distributions of all-, early-, and late-type galaxies are quite similar for the $Z3$ and $Z3B$ samples, implying that a direct comparison between the two samples is justified.

The un-normalized ${\rm d}n/{\rm d}z$ is used to determine the number of galaxies within a volume element defined by $z + {\rm d}z$.  In order to determine $n_g$, we integrate over the entire redshift range with weights given by (${\rm d}n/{\rm d}z$)$^2$:
\begin{equation}
n_g =\int {\rm d}z  \frac{H(z)}{4\pi f_{obs} x^2(z) c}\frac{{\rm d}n}{{\rm d}z}\times \left(\frac{{\rm d}n}{{\rm d}z}\right)^2 ~  /  ~ \int {\rm d}z\left(\frac{{\rm d}n}{{\rm d}z}\right)^2
\end{equation}
where $f_{obs}$ is the observed fraction of the sky and is 0.131 for our masked DR5 sample.  By calculating $n_g$ in this manner, we account for the non-negligible photometric redshift errors, which make the total volume occupied by the galaxies in each of our samples larger than that of a truly volume limited sample.  For each of our two main samples, we use this formalism to measure $n_g$ and compare this to the model $n_g$ in Equation \ref{eq:ng} to determine the value of $M_{cut}$ given $M_0$, $\alpha$, and $\sigma_{cut}$.

When modeling the HOD of early- and late-type galaxies, we are constrained by the fact that the HOD model fraction of late-types, $f_{late}$, must match the observed fraction.  The HOD model fraction of late-types is determined via
\begin{equation}
f_{late} = \frac{1}{n_g}\int {\rm d}Mn(M)\left[f_c(M)N_c(M) + f_s(M)N_c(M)N_s(M)\right]
\end{equation}
In this way, we determine $f_{c0}$ for each given $f_{s0}$, $\sigma_{cen}$, and $\sigma_{sat}$ such that the model $f_{late}$ matches the observed $f_{late}$.  

\section{Measurement Techniques}
\label{sec:meas}
We calculate the angular 2-point correlation function, $\omega(\theta)$, of galaxies using the \cite{LZ} estimator:
\begin{equation}
\omega(\theta) = \frac{DD(\theta) - 2DR(\theta) + RR(\theta)}{RR(\theta)}
\end{equation}
where $DD$ (in our case) is the number of galaxy pairs, $DR$ the number of galaxy-random pairs, and $RR$ the number of random pairs, all separated by an angular distance $\theta \pm \Delta\theta$.  We also calculate angular 2-point cross-correlation functions, for which we also employ the \cite{LZ} estimator:
\begin{equation}
\omega_{{el}}(\theta) = \frac{D_1D_2(\theta) - D_1R(\theta) - D_2R(\theta) + RR(\theta)}{RR(\theta)}
\end{equation}
where $D_1$ and $D_2$ represent the two data samples that are being cross-correlated (note that the single random file can be used in our case since all of our samples have identical angular selections and that we use the subscript $el$ because we will exclusively be cross-correlating early- and late-type galaxies).  In every case, we mask our data and randoms by using the same pixelized mask of R07.

\subsection{Errors and Covariance}
\label{sec:err}
We compute errors and covariance matrices using a method that estimates the statistical error associated with our angular selection and another that estimates the statistical error associated with our radial selection.  We use a jackknife method (e.g., \citealt{Scr02}), with inverse-variance weighting for both errors (e.g., \citealt{Mye05,Mye06}) and covariance (e.g., \citealt{Mye07}) to account for the errors based on our angular selection;  the method is nearly identical to the method described in detail in R07.  The jackknife method works by creating many subsamples of the entire data set, each with a small part of the total area removed.  We found in R07 that 20 jack-knife subsamplings are sufficient to create a stable covariance matrix.  These 20 subsamples are created by extracting a contiguous grouping of 1/20th of the unmasked pixels in 20 separate areas.  Our covariance matrix, $C_{\rm jack}$, is thus given by
\begin{equation}
\begin{array}{ll}
C_{i,j, {\rm jack} }=  C_{\rm jack}(\theta_i,\theta_j) & ~\\
= \frac{19}{20} \sum_{k=1}^{20}[\omega_{full}(\theta_i) - \omega_{k}(\theta_i)][\omega_{full}(\theta_j) - \omega_{k}(\theta_j)] & ~
\end{array}
\label{eq:JK}
\end{equation} 
\noindent where $\omega_{k}(\theta)$ is the value for the correlation measurement omitting the $k$th subsample of data and $i$ and $j$ refer to the $i^{th}$ and $j^{th}$ angular bin. The jackknife errors are simply the square-root of diagonal elements of the covariance matrix.  

We must use a separate method to account for uncertainties introduced by our radial selection.  In essence, we are attempting to measure the auto-correlation functions of galaxies for a given redshift distribution (since the redshift distribution figures prominently in our models).  Our defined cuts on photometric redshift  do not uniquely produce the redshift distributions displayed in Figure \ref{fig:dndz}.  In order to account for this, we re-sample the photometric redshift catalog to create `perturbed' samples whose redshift distributions match those of the original sample.  We take the redshift of each galaxy to be randomly selected from its PDF and re-calculate $M_r$ based on this redshift.  If these perturbed redshifts and magnitudes satisfy our selection criteria, they are included in the new sample of galaxies.  In order to adequately reproduce the redshift distributions of Figure \ref{fig:dndz}, we find that we can only allow galaxies with $z < 0.32$ into our perturbed sample.  For the early- and late-type galaxies, we also perturb the type value based on the type-error (assuming it is Gaussian) when producing our perturbed samples.  The type errors scale linearly with the photometric redshifts errors.  Thus we use
\begin{equation}
t_n = t + (z_n - z)\sigma_t/\sigma_z
\end{equation}
where $t$ is the galaxy type, $t_n$ and $z_n$ are the perturbed type and redshift, and $\sigma_t$ and $\sigma_z$ are the type error and photometric redshift error of each object obtained from the DR5 {\tt PhotoZ} table. 

For each of our galaxy samples, we create ten perturbed samples.  The percentage of galaxies that match between samples varies between 77\% and 85\% for any given parent sample (variation within any group of ten perturbed samples is less than 1\%, e.g., the percentage of matching galaxies is always between 76.5\% and 77.4\% for late-type galaxies from the $Z3B$ sample).  We calculate $\omega(\theta)$ for each of the perturbed samples and calculate $C_{\rm z}$:
\begin{equation}
\begin{array}{ll}
C_{i,j, {\rm z} }= C_{\rm z}(\theta_i,\theta_j) & ~ \\
 =  \sum_{k=1}^{10}f_m[\omega_{ave}(\theta_i) - \omega_{k}(\theta_i)][\omega_{ave}(\theta_j) - \omega_{k}(\theta_j)] & ~
\label{eq:JK}
\end{array}
\end{equation}
where $\omega_{ave}$ is the average auto-correlation of each of the ten perturbed samples and $f_m$ is the average fraction of galaxies that match between sample $k$ and the other nine samples.  We find that typically $C_{i,j, {\rm z} }\sim 0.5 C_{i,j, {\rm jack} }$, meaning that they are small but non-negligible.

We thus combine $C_{\rm z}$ and $C_{\rm jack}$ to obtain the full covariance matrix for each sample (i.e. $C_{i,j} = C_{i,j, {\rm jack}} + C_{i,j, {\rm z}}$).  To properly constrain fit parameters, we minimize the $\chi^2$ using our covariance matrixes via the equation
\begin{equation}
\chi^2 = \sum_{i,j}[\omega(\theta_i) - \omega_{m}(\theta_i)]C_{i,j}^{-1}[\omega(\theta_j) - \omega_{m}(\theta_j)]
\label{eq:Chi}
\end{equation} 
where $\omega_m(\theta)$ refers to the model angular 2-point correlation function.

\section{Measurements}
\label{sec:res}
\begin{figure*}
\centering
\begin{minipage}{140mm}
\includegraphics[width=140mm]{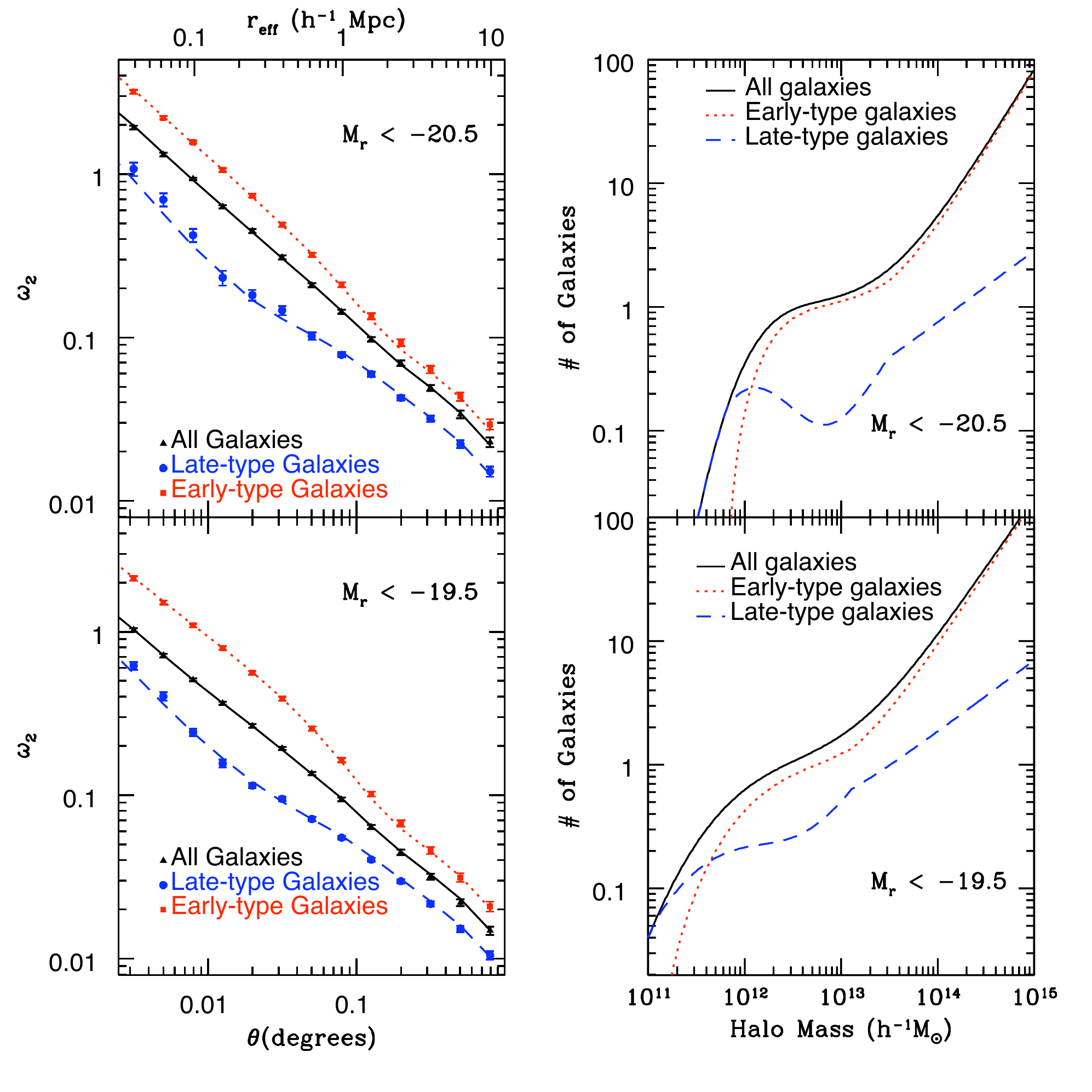}
\caption{The left panels display the measured angular auto-correlation functions for galaxies with $M_r < -20.5$ (top) and $M_r < -19.5$ (bottom), for all (black triangles) early- (red squares) and late-type (blue circles) with lines that correspond to the best-fit model (black solid for all, red dotted for early-, and blue dashed for late-type).  The right panels display the best-fit HOD for all (black solid), early- (red dotted), and late-(blue dashed) type galaxies for $M_r < -20.5$ (top) and $M_r < -19.5$ (bottom).}
\label{fig:z3VL}
\end{minipage}
\end{figure*}

We have measured the angular 2-point correlation functions and found the best-fit HOD for galaxies in two luminosity threshold samples; $Z3$ ($z < 0.3$ and $M_r < -19.5$), and $Z3B$ ($z < 0.3$ and $M_r < -20.5$).  For each sample we also found the parameters that best-fit the $\omega(\theta)$ of early- and late-type galaxies.  The values of the best-fit HOD parameters for each full sample can be found in Table \ref{tab:HOD}, and the best-fit parameters for the early- and late-type samples can be found in Table \ref{tab:HODel}.

The top-left panel of Figure \ref{fig:z3VL} displays the measured $\omega(\theta)$ for galaxies in the $Z3B$ sample for all (black triangles), early- (red triangles), and late- (blue triangle) type galaxies with the best-fit model $\omega(\theta)$ plotted with correspondingly colored lines (solid for all, dashed for late-, and dotted for early-type galaxies).  The fit to all galaxies is excellent, as fitting between 0.003$^\deg$ and 1$^\deg$ (10 degrees of freedom) yields $\chi^2 = 3.9$.  The fit to the early- and late-type galaxies is tolerable, as for the 23 degrees of freedom $\chi^2 = 19.2$.  The model performs slightly better for the early-type galaxies ($\chi^2$ = 8.4), than for the late-types ($\chi^2 = 10.8$).  The largest discrepancies are at small scales, where the late-type model is too low.  We note that if we had assumed that early- and late-type galaxies mix freely within halos, the fits would have been significantly worse, as we find the minimum $\chi^2$ is 55 for such a model when fitting the measurements over the same angular range.

\begin{table*}
\centering
\begin{minipage}{10in}
\caption{The best-fit values of the HOD parameters and the associated $\chi^2$ values for the two main samples studied.  All masses are in units $M_\odot h^{-1}$.} 
\begin{tabular}{lccccccccc}
\hline
\hline
Sample & $\alpha$ & ${\rm log}\left(M_{cut}\right)$ &  ${\rm log}\left(M_{0}\right)$  & $\sigma_{cut}$ & $\chi^2/{\rm dof}$ & ${\rm log}\left(M_{eff}\right)$ &  $b_1$ & $n_g$ $(h^3 {\rm Mpc}^{-3})$ & $f_{sat}$\\
\hline
$Z3$ & 1.14$^{+0.02}_{-0.01}$ & 11.866 & 13.11$\pm$0.01 & 0.7$^{+0.06}_{-0.09}$ &  1.9/10 & 13.13 & 1.09  & 0.0102 & 0.148\\
$Z3B$ & 1.268$^{+0.026}_{-0.024}$ & 12.115 & 13.488$^{+0.009}_{-0.011}$ & 0.41 $^{+0.13}_{-0.14}$ & 3.9/10 & 13.21 & 1.17 & 0.0041& 0.130\\
\hline
\label{tab:HOD}
\end{tabular}
\end{minipage}
\end{table*}


\begin{table*}
\centering
\begin{minipage}{10in}
\caption{The best-fit values of the HOD parameters and the associated $\chi^2$ values for the early- and late-type samples studied.} 
\begin{tabular}{lcccccccccc}
\hline
\hline
Sample  & $f_{c0}$ &  $f_{s0}$&  $\sigma_{cen}$ &   $\sigma_{sat}$&   $\chi^2/{\rm dof}$ &  $b_{1, late}$ &   $b_{1, early}$ & $f_{late}$ & $f_{sat,late}$ & $f_{sat,early}$\\
\hline
$Z3$ &  0.38 & 0.56$^{+0.04}_{-0.02}$ & 0.84$\pm$0.09  & 0.80$^{+0.09}_{-0.06}$& 15.7/23 &  0.89 & 1.16 & 0.498 & 0.118 & 0.180\\
$Z3B$ &  0.437 & 0.38$\pm$0.02 & 0.63$^{+0.08}_{-0.10}$ & 0.35 $\pm$0.05 & 19.2/23  & 0.98 & 1.27& 0.370 & 0.149 & 0.118\\
\hline
\label{tab:HODel}
\end{tabular}
\end{minipage}
\end{table*}

\begin{figure}
\includegraphics[width=84mm]{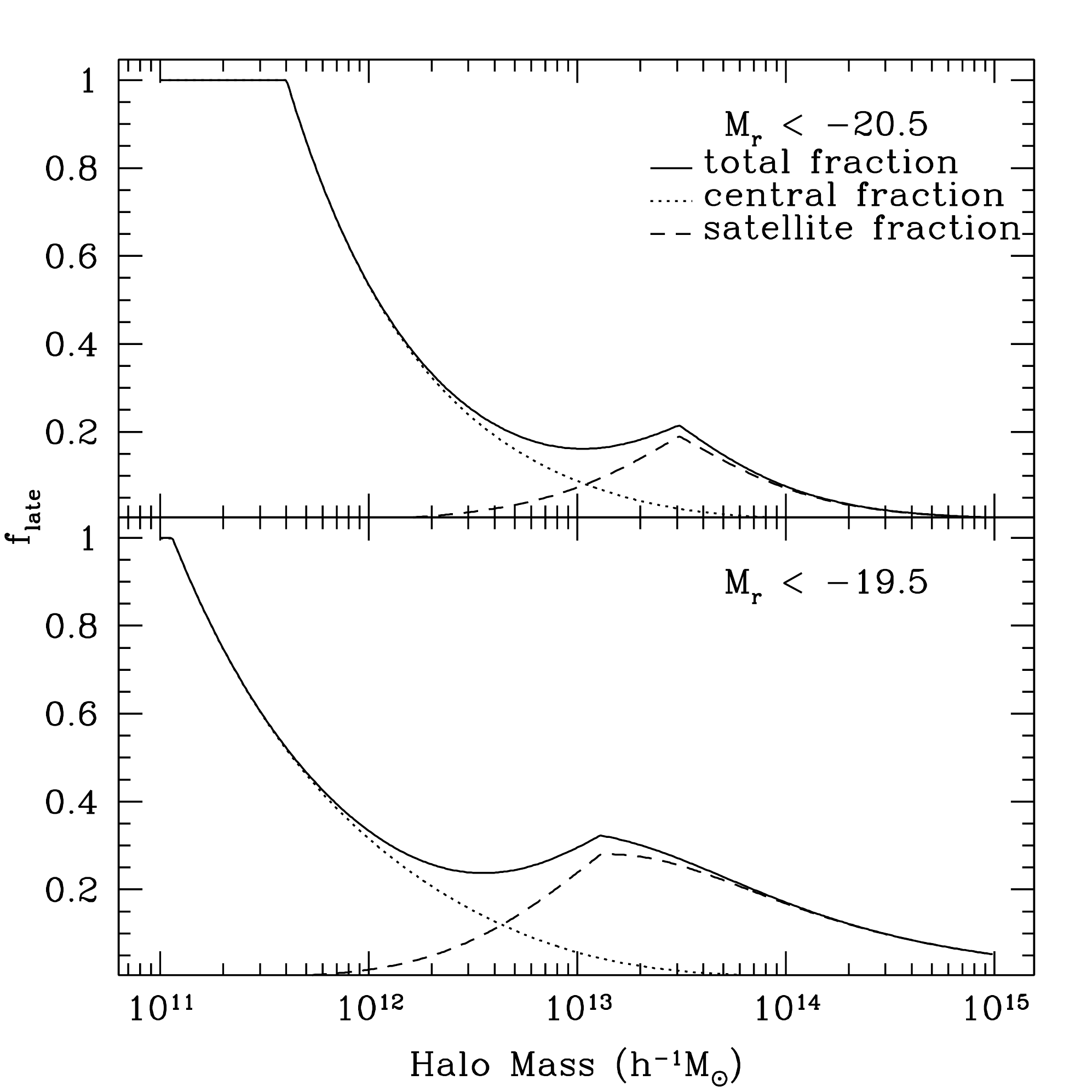}
\caption{The fraction of galaxies that are late-type as a function of halo-mass.  The top panel displays the information for galaxies with $M_r < -20.5$ and the bottom panel for galaxies with $M_r < -19.5$.  In both panels, the full fraction is displayed with a solid line, the central fraction is displayed with a dotted line, and the satellite fraction is displayed with a dashed line.  Note that the information is displayed such that the satellite fraction and central fraction add to the full fraction.}
\label{fig:latefrac}
\end{figure}

The top-right panel of Figure \ref{fig:z3VL} displays the best-fit HOD for all (black), early- (red), and late- (blue) type galaxies from the $Z3B$ sample.  The HOD for all galaxies shows inflection points around $M_{cut} = 1.30 \times 10^{12}h^{-1}M_{\odot}$, which defines the mass scale at which halos host a central galaxy, and $M_0 = 3.08 \times 10^{13}h^{-1}M_{\odot}$, which defines the mass scale at which halos will host satellite galaxies.  The best-fit late-type HOD shows a local minimum at close to $10^{13}h^{-1}M_{\odot}$.  This shape is a consequence of the model --- the fraction of central late-type galaxies decreases as the mass increases and thus the late-type HOD decreases until the halos are massive enough to host satellite galaxies.  Even so, the slope of the late-type HOD is significantly smaller than for the overall HOD, allowing the fraction of late-type galaxies to be the largest in small mass halos and smallest in high mass halos.  This can be seen clearly in the top panel of Figure \ref{fig:latefrac}, where the fraction of late-type galaxies is plotted against halo mass.  The decrease is nearly monotonic except for a feature with a local maximum right at $M_0$.  This is consistent with the density-morphology relation \citep{Dre80}, as the fraction of late-type galaxies decreases as halo mass increases.  The model allows $f_c$ to increase from 0.38 at $M = M_{cut}$ to 1 near $3 \times 10^{11}h^{-1}M_{\odot}$.  We note that we also tried a model where $f_c = f_{c0}$ for $M < M_{cut}$, but we did not obtain acceptable fits.  


The measured and best-fit model $\omega(\theta)$ for galaxies from $Z3$ are plotted in the bottom-left panel of Figure \ref{fig:z3VL}.  The best-fit model to the entire sample is acceptable, as $\chi^2 = 1.9$ fitting between 0.003$^\deg$ and 1$^\deg$ (10 degrees of freedom).  The model fits for the early- and late-type galaxies also perform well, as the total $\chi^2 = 15.7$ (23 degrees of freedom).  Once again, the fit is slightly better for early-type galaxies ($\chi^2 = 6.1$) than for late-type galaxies ($\chi^2 =$ 9.6). The largest disagreements are at small scales where the late-type model is not quite large enough.  Once more, we note that our model out-performs one in which we allow early- and late-type galaxies to mix freely within halos; we find the minimum $\chi^2$ for this class of model is 68 when fitting over the same angular range.  

The galaxies in $Z3B$ form a brighter subset of the $Z3$ galaxies.  Thus, as can be seen in the righthand panels of Figure \ref{fig:z3VL}, the HOD of each $Z3$ sample is larger than its $Z3B$ counterpart at every mass scale (though they are all nearly equal at $10^{12}h^{-1}M_{\odot}$).  The inflection points of the $Z3$ HOD for the full sample (occuring around $M_{cut} = 11.73 \times 10^{11}h^{-1}M_{\odot}$ and $M_0 = 1.29 \times 10^{13}h^{-1}M_{\odot}$) are not as pronounced as for the $Z3B$ sample, which is due to the larger value of $\sigma_{cut} = 0.7$ (as compared to the $\sigma_{cut} = 0.4$ for the $Z3B$ sample).  The best-fit HOD of the late-type galaxies shows similar behavior to the $Z3B$ late-type HOD, as once more the late-type fraction is greatest at small halo masses and least at large halo masses.  The overall number of late-type galaxies is more than four times higher, and this very nearly matches the difference in the late-type HODs for $M_{halo} > 10^{13}h^{-1}M_{\odot}$.  At smaller halo masses, there are more significant differences.  The $Z3$ late-type HOD does not have a local minimum, only a significant inflection point.  This is primarily due to the fact that the parameter which governs the decrease of $f_c$, $\sigma_{cen}$, has increased from 0.63 to 0.82.  

The bottom panel of Figure \ref{fig:latefrac} displays the fraction of late-type galaxies for the best-fit HOD model of the $Z3$ sample, as a function of halo mass.  Again, there is a nearly monotonic decrease with a local maximum at $M_1$.  As plotted with dashed lines, this local maximum is due to a peak in the late-type satellite fraction.  This peak exists at $M_0$ because, at smaller halos masses, the total fraction of galaxies that are satellites drops sharply and the fraction of galaxies that are late-type satellites must as well.  (Note that the data is displayed such that the satellite and central fractions add to the total fraction.)  

The late-type satellite fraction is larger for the $Z3$ sample, but this due primarily to the fact that the overall late-type fraction has increased from 0.37 to 0.498.  The overall fraction of late-type galaxies that are satellites is smaller for the $Z3$ sample (as presented in Table 2, it is 0.118 for $Z3$ and 0.149 for $Z3B$).  This is due to the fact that the bulk of the late-type galaxies are central galaxies in low mass halos, and the total fraction of centrals is higher because the number density of halos is larger at small mass.  Conversely, the satellite fraction of early-type goes up for the $Z3$ sample.  Essentially, the model implies that a majority of the late-type galaxies with $-19.5 < M_r < -20.5$ are central galaxies occupying low-mass halos and the majority of the early-type galaxies with $-19.5 < M_r < -20.5$ are satellites in higher-mass halos.

\begin{figure}
\includegraphics[width=84mm]{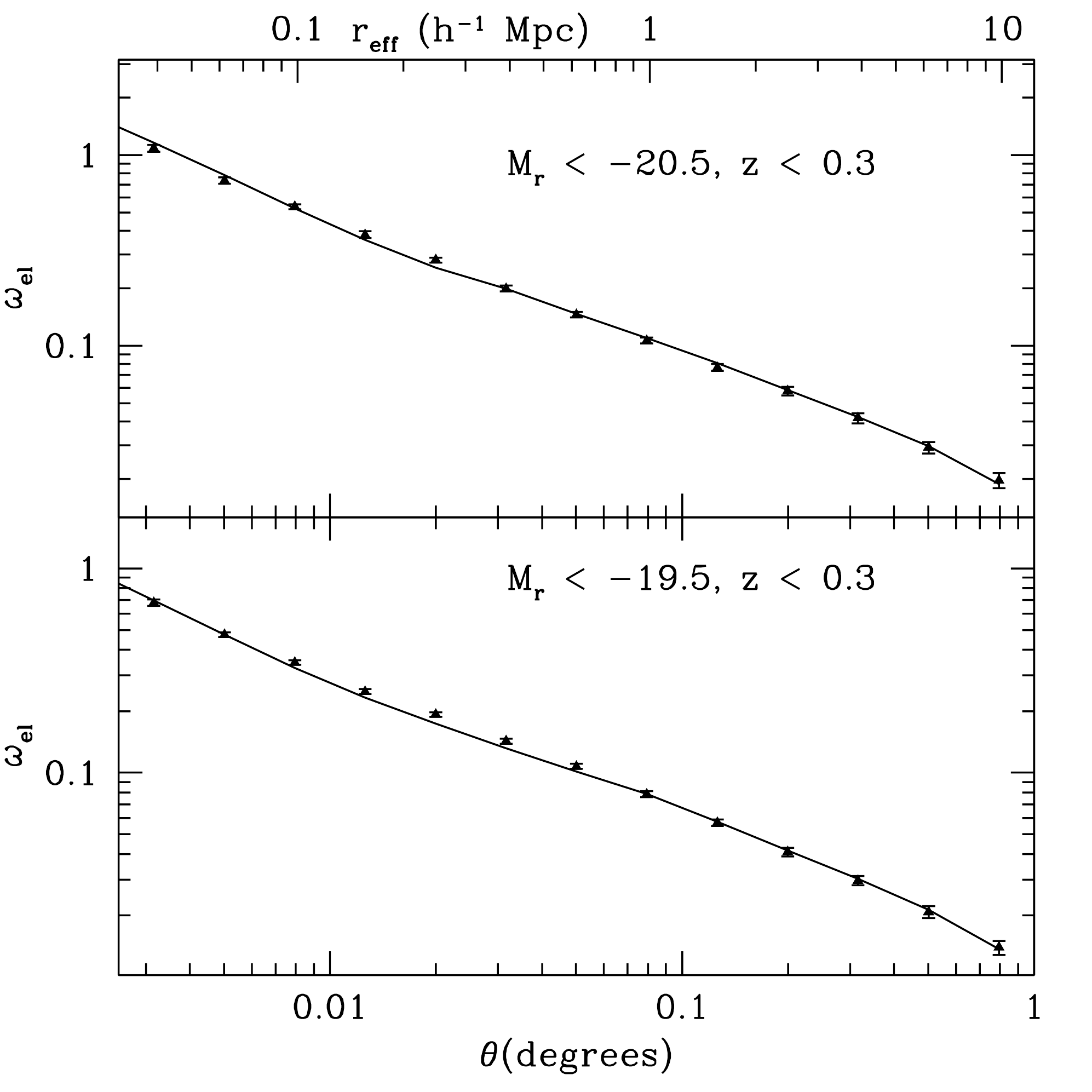}
\caption{The measured cross-correlations of early- and late-type galaxies in for the $Z3$ sample, volume limited with $z < 0.3$ and $M_r < -19.5$ (bottom), $Z3B$ sample, volume limited with $z < 0.3$ and $M_r < -20.5$ (top) are displayed in black triangles.  The appropriate model cross-correlation, determined using the best-fit HOD of the early- and late-type autocorrelations, is displayed with a solid black line in panel. }
\label{fig:z4scross}
\end{figure}

We also measure the 2-point cross-correlation function of early- and late-type galaxies for each sample.  These measurements are plotted in Figure \ref{fig:z4scross} (black triangles) along with the model $\omega_{el}(\theta)$ that results from using the best-fit parameters determined from the autocorrelation measurements.  The models appear close to the measurements, and the $\chi^2$ are 16.1 for $Z3$ and $\chi^2 = 15.3$ for $Z3B$.  These values are impressive considering the size of the error bars and the fact that we did not specifically fit for these measurements.  Once again, the best-fit models which include mixing are significantly worse ($\chi^2 = 92.4$ for $Z3$ and $\chi^2 = 86.6$ for $Z3B$).  Thus, in every example, our model, which segregates early- and late-type galaxies to the maximal extent, performs significantly better than one in which the early- and late-type galaxies are allowed to mix freely.

The model cross-correlation functions match the measurements quite well on large scales, which is what one should expect, as the 2-halo cross-power spectrum is just $\sqrt{P(k,z)_{late}P(k,z)_{early}}$.  The agreement suggests that there are not any major systematic problems in our construction of the redshift distributions of the early- and late-type galaxies.  When we model the angular cross-correlation, effectively there is a term $\int ({\rm d} n/{\rm d}z_{late})({\rm d} n/{\rm d}z_{early})$.  If the true redshift distributions differ greatly from those we estimate, then our models would not be able to simultaneously fit the large scale autocorrelation and cross-correlation function measurements.  



\section{Discussion}
\label{sec:dis}
We have measured the angular auto-correlation functions of galaxies photometrically selected from the SDSS DR5.  We have used these measurements to constrain the HOD of the galaxies and determine its dependence on luminosity, and galaxy type.  We have found that that the fact we are using photometric redshifts requires a special prescription for determining number densities (see \S\ref{sec:dndz}) and introduces an extra source of statistical error (see \S\ref{sec:err}).  Most interestingly, we have found that in order to simultaneously model the clustering of early- and late-type galaxies and their cross-correlation, we cannot allow them to mix freely within halos.

\begin{figure}
\includegraphics[width=84mm]{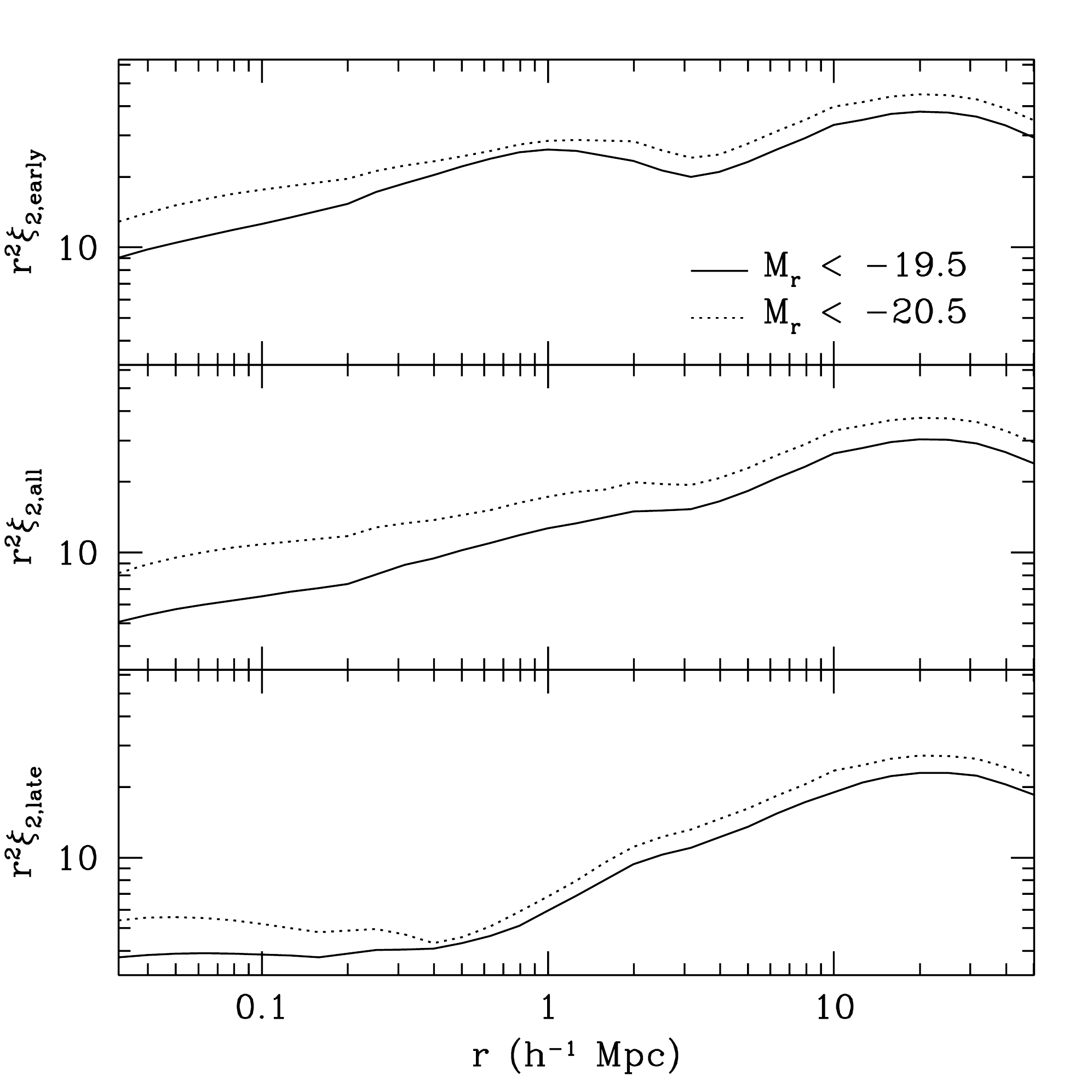}
\caption{The best-fit model real-space 2-point correlation functions multiplied by $r^2$ for early- (top), all (middle), and late- (bottom) type galaxies for the $Z3$ sample, volume limited with $z < 0.3$ and $M_r < -19.5$ (solid lines), $Z3B$ sample, volume limited with $z < 0.3$ and $M_r < -20.5$ (dotted lines).}
\label{fig:ep2all}
\end{figure}

Further insight can be gained by looking at the real-space 2-point correlation functions of our best-fit models.  The $\xi(r)$ of the early-, all, and late-type galaxies are displayed in the top, middle, and bottom panels of Figure \ref{fig:ep2all}, with solid and dotted lines representing the $Z3$ and $Z3B$ samples.  Due to the wide range in scales, we multiply each $\xi$ by $r^2$, which allows the differences between each correlation function to be seen clearly.  The differences between the model $\xi(r)$ of the $Z3$ and $Z3B$ samples are in line with what one would expect.  The samples share the same median redshift ($\bar{z} \sim$ 0.25), so direct comparison is valid.  The model $\xi_{all}(r)$ of $Z3B$ is consistently higher than that of $Z3$, as one would expect given the differences in luminosity.  Interestingly, the $\xi(r)$ of the early- and late-type samples increase by a smaller factor than the full sample.  The $\xi(r)$ amplitudes for the $Z3B$ sample are bolstered not only by the increase in luminosity but also due to the decrease in the fraction of late-type galaxies ($\sim 0.5$ compared to $\sim 0.37$). 

\subsection{Comparison With Previous Results}
We can investigate further by comparing our work with that of Z05, who found the best-fit HOD for galaxies from the spectroscopic portion of the SDSS ($\bar{z} \sim 0.1$).  The model used by Z05 also did not include a $\sigma_{cut}$ parameter.  \cite{Tink08} constrained the HOD of SDSS galaxies using a HOD model that does include $\sigma_{cut}$, but the facts that Z05 constrain galaxy samples with luminosity thresholds that match ours ($M_r< -19.5$ and $M_r< -20.5$) and that the results of \cite{Tink08} generally agree with Z05, make it a more appropriate reference.  

\begin{figure}
\includegraphics[width=84mm]{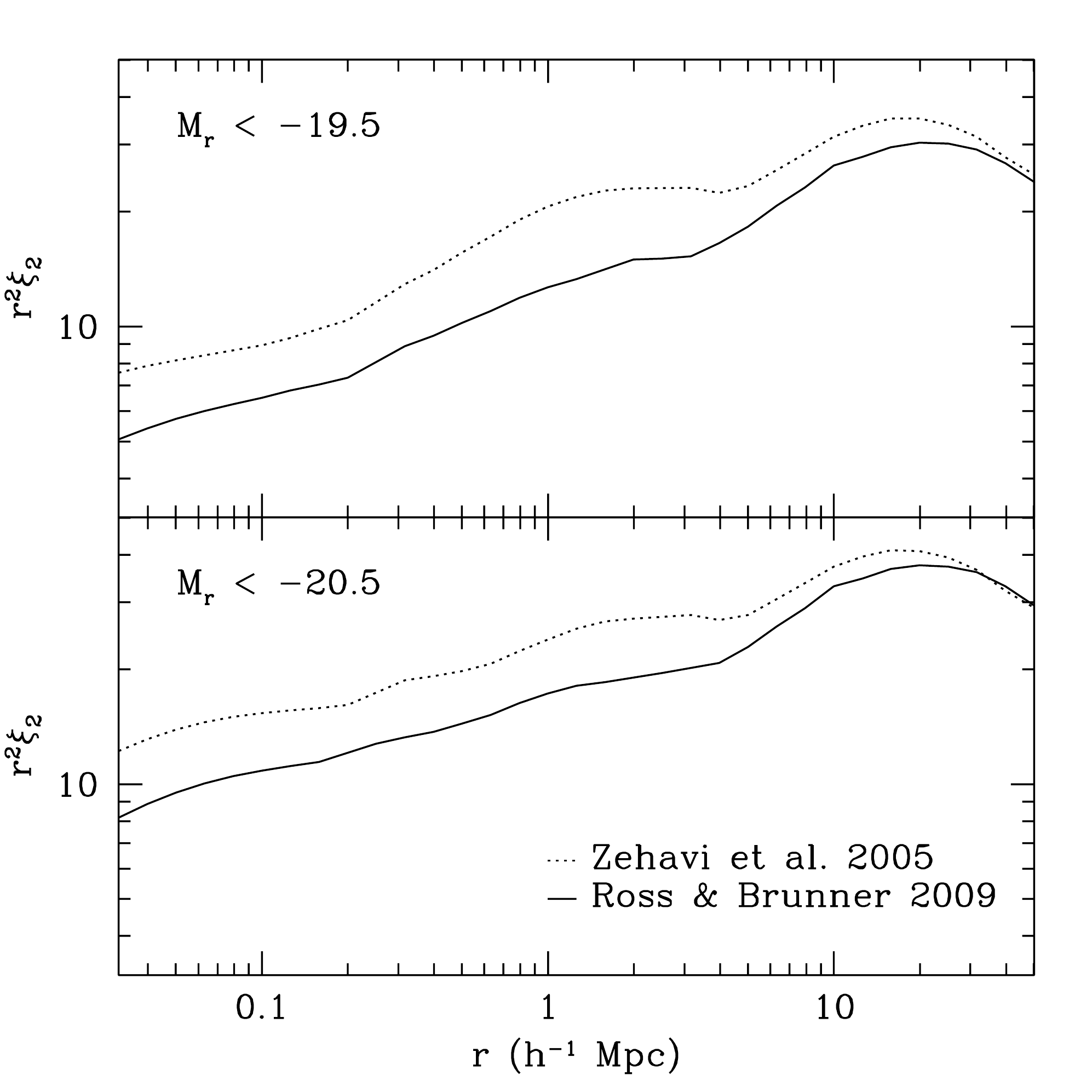}
\caption{The top panel displays the best-fit model real-space 2-point correlation functions multiplied by $r^2$ for our data volume limited with $z < 0.3$ and $M_r < -19.5$ (solid lines) compared to the best-fit model real-space 2-point correlation functions from Z05 with $M_r < -19.5$ (dashed lines) and the same model shifted to $z = 0.25$ assuming passive evolution (dotted lines).  The bottom panel displays the same information for samples with $M_r < -20.5$. }
\label{fig:zehcom}
\end{figure}

Despite the differences in the samples used and in the models, the differences between the Z05 best-fit model parameters and ours can be explained by the fact that Z05 used a fairly different cosmological model.  In order to compare our results to those of Z05, we calculate real-space 2-point correlation functions using the  Z05 cosmological model ($\Omega_{total} = 1$, $\Omega_m = 0.3$, $h=0.7$, $\Gamma = 0.21$, $\sigma_8 = 0.9$) and best-fit HOD parameters.  In Figure \ref{fig:zehcom} we display our best-fit real-space 2-point correlation functions for our $Z3$ and $Z3B$ samples (solid lines) compared to the Z05 best-fit real-space 2-point correlation functions for $M_r < -19.5$ and $M_r < -20.5$ (dotted lines).  Once again we multiply each $\xi$ by $r^2$, in order to clearly see the differences between the respective correlation functions.  We multiply the Z05 correlation amplitudes by a factor that allows for passive evolution between $z = 0.25$ and $z = 0.1$ given by (see, e.g., \citealt{Wake08})
\begin{equation}
\frac{\xi_{hi}(r)}{\xi_{lo}(r)} = \left(\frac{b_{lo}-1 + D_{hi}/D_{lo}}{b_{lo}}\right)
\label{eq:pev}
\end{equation}
where the subscripts $lo$ and $hi$ refer to the appropriate factors at the lower and higher redshifts and $D$ is the linear growth factor (see, e.g., \citealt{MW96}).  We note that applying this factor simply allows for a proper comparison between the two clustering signals --- we are not arguing that galaxies with $M_r < -19.5$ should passively evolve between z = 0.25 and 0.1.  The comparison is particularly apt for the $M_r < -19.5$ samples, as the co-moving number densities are the same to 3 significant figures (0.0102 $h^3$Mpc$^{-3}$), and slightly less so for the $M_r < -20.5$ sample as the number density in our sample is roughly 33\% higher than the Z05 sample.  

For both samples, the correlation functions have similar amplitudes at large scales, but the Z05 amplitudes are significantly larger at small scales.  This is what we would generally expect, as merging halos between 0.25 and 0.1 should increase the overall satellite fraction and thus increase the amplitude of the one halo term in the correlation function.  This is indeed the case in the best-fit models. As presented in \cite{ZZ07}, the satellite fraction is $\sim$0.2 for $L_{*}$ galaxies from the SDSS spectroscopic, while ours is $\sim$0.15.  The decrease is perhaps slightly more than one would expect between $z\sim0.1$ and $z\sim0.25$, but the general trend is as expected.  We therefore do not find any significant disagreement between the clustering of galaxies in our photometric samples and that of galaxies from the SDSS spectroscopic sample.



On the other hand, our results differ from those of Z05 with respect to the early- and late-type galaxies.  The manner in which Z05 separated galaxies into different samples differs slightly from ours, as Z05 uses a $K$-corrected color cut and we use an estimated spectral type.  One would expect the galaxies Z05 classify as ``blue" to predominantly be late-type (and the ones Z05 classify as ``red" to predominantly be early-type).  It is thus somewhat surprising that the Z05 measurements were well-fit by a model allowing red and blue galaxies to mix freely within halos --- while we find, that in all cases, a model that separates the early- and late-type galaxies to the maximal extent possible (given the statistics) provides a much better fit to our data.  The discrepancy is likely due to the fact that the smallest scales are most strongly affected by this treatment of the galaxies and that we probe significantly smaller scales than Z05 (our fits extend to $r_{eq} \sim 0.03~ h^{-1}$ Mpc).

We can also easily compare our results to R07, as we use the same volume limited samples and early-/late-type splits.  Comparing the $b_1$ values of the full samples, they are quite close, but slightly inconsistent given the 1$\sigma$ errors quoted in R07.  This is due to the fact that in R07 the $b_1$ values were measured directly from the data, while the $b_1$ values quoted in this work are from the models derived from the best-fit HOD.  Therefore, one would expect slight disagreement.  The disagreement is never greater that 4\%, so this is not worrisome.  The disagreement is greater for the early- and late-type samples, which is likely due to the treatment of the photometric redshift distributions.  R07 determined a model $\omega_{2,DM}$ using the redshift distribution for all of the galaxies in each respective sample and used it to find the all, early-, and late-type $b_1$ parameters for the sample.  As can be seen in Figure \ref{fig:dndz}, the redshift distributions of the early- and late-type galaxies show different shapes.  In general, the early-type distributions are more narrow, and the late-type distributions wider than the full distribution.  Thus, using the full sample to find $b_1$ of late-type galaxies would cause the parameter to be under-estimated, and using it to find $b_1$ of early-type galaxies would cause the parameter to be overestimated.  This appears to be the case for R07, as the $b_1$ of the late-type galaxies are consistently smaller than ours and the early-type $b_1$ are consistently higher.

\subsection{Mixing}
Our best-fit models of early- and late-type galaxy clustering, constrained via our auto-correlation measurements, provide good fits to the respective auto-correlation measurements and also acceptable fits to the cross-correlation measurements (see Figure \ref{fig:z4scross}).  We have investigated changing other parameters in the models to determine if there is an alternative course to separating early- and late-type galaxies into different halos, but none have provided acceptable fits.  For example, we have tried a wide range of models for the density profiles of the late-type galaxies within halos that are different from the standard NFW (though always spherically symmetric), but we were not able to significantly improve the model fits (either while allowing mixing or not).  We have also tried using different forms for the concentration parameter of late-type galaxies, but again, this produced no meaningful improvement.  Given the data at hand, we are, therefore, convinced that our model represents the optimal way of modeling the clustering of early- and late-type galaxies.


By looking at the measured $\omega(\theta)$, one can see why models with mixing can not reproduce our measurements.  For any model that allows even mixing, if a model for late-type galaxies is forced to be closer to the model for all galaxies, the early-type model will have to do the same (in the case where the number densities of early- and late-type galaxies are equal, the response should be entirely symmetric).  Our measurements for both galaxy samples show the measured $\omega(\theta)$ of late-types getting closer to the measured $\omega(\theta)$ of all galaxies as the scale gets smaller, while the early-type measurements do not get any closer to the measurements for all galaxies.  Segregating the galaxies allows more freedom in each galaxy-type's model $\omega(\theta)$ relative to the model for all of the galaxies.  This can be illustrated by imagining two samples that are completely segregated but have identical clustering properties within halos.  In combining these two samples, the number of close pairs will only double (along with the total number of objects) and the correlation function for the entire sample will be half as large as for either of the original samples.  In the case where the two samples are mixed within halos, combining the two samples would quadruple the number of close pairs, thus producing the same result for the correlation function.

The shapes of the model $\omega_{el}(\theta)$ elucidate the minimum degree to which Equations \ref{eq:cent} and \ref{eq:sat} require mixing of galaxy types.  If, for example, the galaxy types were allowed to be completely segregated, the model cross-power spectrum one halo term would be zero and the model cross-correlation would be completely flat at small scales.  Clearly, the best-fit models are not allowing such extreme segregation.  In most cases, Equations \ref{eq:cent} and \ref{eq:sat} will require that many late-type galaxies are satellites of central early-type galaxies, who also have many early-type satellites (this will happen for any halo mass where $f_s$ is greater than $f_c$).  

\section{Conclusions}
\label{sec:con}
We have measured the angular 2-point correlation functions of galaxies drawn from volume limited samples of SDSS DR5 galaxies with $z < 0.3, M_r < -19.5$ and $z < 0.3, M_r < -20.5$,   each of which are further subdivided into early- and late-type galaxy subsamples.  By modeling the angular 2-point correlation function, we have shown, for the first time, that the best halo model is one in which early- and late-type galaxies are segregated to the maximal extent possible.  Previous studies (such as Z05) modeled the clustering of red and blue galaxies (which should predominantly be early- and late-type galaxies, respectively) by allowing mixing between the galaxy types within halos; these studies, however, did not probe to the same small scales we have, which is where the models that allow mixing disagree the strongest with our measurements.

We plan to follow-up this work by using data from the SDSS DR7 to constrain the HOD as a function of redshift.  The analysis techniques presented in this work provide a foundation upon which to base this extension and the improved photometric redshifts of the DR7 data should enable a reliable determination of the evolution of clustering as a function of galaxy type to $z < 0.4$. 

\section*{Acknowledgements}
A.J.R and R.J.B acknowledge support for Microsoft Research, the University of Illinois, and NASA through grant NNG06GH156.  The authors made extensive use of the storage and computing facilities at the National Center for Super Computing Applications and thank the technical staff for their assistance in enabling this work.

We thank Ani Thakar and Jan Van den Berg for help with obtaining a copy of the SDSS DR5 databases.  We thank Adam D. Myers and David Wake for helpful discussions and comments that improved this work.  We thank an anonymous referee for (quite timely) comments that improved both our modeling and the clarity of its written presentation.

\label{lastpage}

\begin{thebibliography}{7}
\bibitem[Abazajian et al.(2005)]{Ab05} Abazajian, K., et al.\ 2005, AJ, 129, 1755
\bibitem[Ball et al.(2008)]{Ball08} Ball, N.~M., Loveday, J., \& Brunner, R.~J.\ 2008, MNRAS, 383, 907
\bibitem[Blake et al.(2008)]{Blake} Blake, C., Collister, A., Lahav, O.\ 2008, MNRAS, 385, 1257 
\bibitem[Budav{\'a}ri et al.(2003)]{Bud03} Budav{\'a}ri, T., 
et al.\ 2003, ApJ, 595, 59 
\bibitem[Bullock et al.(2001)]{Bull01} Bullock, J.~S., Kolatt, T.~S., Sigad, Y., Somerville, R.~S., Kravtsov, A.~V., Klypin, A.~A., Primack, J.~R., \& Dekel, A.\ 2001, MNRAS, 321, 559
\bibitem[Cooray \& Sheth(2002)]{CooSh02} Cooray, A., \& Sheth, R.\ 2002, Phys. Rep., 372, 1
\bibitem[Cowie et al.(1996)]{Cow96} Cowie, L.~L., Songaila, A., Hu, E.~M., \& Cohen, J.~G.\ 1996, AJ, 112, 839 
\bibitem[Croton et al.(2006)]{Cr06} Croton, D.~J., Norberg, P., Gazta{\~n}aga, E., \& Baugh, C.~M.\ 2007, MNRAS, 379, 1562  
\bibitem[Dressler(1980)]{Dre80} Dressler,~A., 1980, ApJ, 236, 351
\bibitem[Dressler et al.(1997)]{Dress97} Dressler, A., et al.\ 
1997, ApJ, 490, 577 
\bibitem[Fukugita et al.(1996)]{F} Fukugita, M., 
Ichikawa, T., Gunn, J.~E., Doi, M., Shimasaku, K., \& Schneider, D.~P.\ 
1996, AJ, 111, 1748
\bibitem[Gunn et al.(1998)]{C} Gunn, J.~E., et al.\ 1998, 
AJ, 116, 3040
\bibitem[Jain et al.(2003)]{jain03} Jain, B., 
Scranton, R., \& Sheth, R. K.\ 2003, MNRAS, 345, 62
\bibitem[Kauffmann et al.(1997)]{Kauf97} Kauffmann, G., 
Nusser, A., \& Steinmetz, M.\ 1997, MNRAS, 286, 795
\bibitem[Landy \& Szalay(1993)]{LZ} Landy S. D., Szalay A. S., 1993, ApJ, 412, 64
\bibitem[Limber(1954)]{lim} Limber, D. N. 1954, ApJ, 119, 655
\bibitem[Lupton et al.(2002)]{L} Lupton, R.~H., Ivezic, Z., Gunn, J.~E., Knapp, G., Strauss, M.~A., \& Yasuda, N.\ 2002, Proc. SPIE, 
4836, 350
\bibitem[Madgwick et al.(2003)]{Ma03} Madgwick, D.~S., et 
al.\ 2003, MNRAS, 344, 847 
\bibitem[Mo \& White (1996)]{MW96} Mo, H.~J., White, S.~D., 1996, MNRAS, 282, 347
\bibitem[Myers et al.(2005)]{Mye05} Myers,~A.~D., Outram,~P.~J., Shanks,~T., Boyle,~B.~J., Croom,~S.~M., Loaring,~N.~S., Miller,~L., \& Smith,~R.~J. 2005, MNRAS, 359, 741 
\bibitem[Myers et al.(2006)]{Mye06} Myers, A.~D., et al.\ 
2006, ApJ, 638, 622
\bibitem[Myers et al.(2007)]{Mye07} Myers, A.~D., et al.\ 
2007, ApJ, 658, 85 
\bibitem[Navarro, Frenk, \& White(1997)]{NFW} Navarro J. F., Frenk C. S., White S. D. M., 1997, ApJ, 490, 493
\bibitem[Nishimichi et al.(2006)]{Nish06} Nishimichi, T., Kayo, I., Hikage, C., Yahata, K., Taruya, A., Jing, Y.~P., Sheth, R.~K., \& Suto, Y.\ 2007, PASJ, 59, 93 
\bibitem[Hopkins(2004)]{Hop04} Hopkins, A.~M.\ 2004, ApJ, 
615, 209
\bibitem[Norberg et al.(2001)]{Norberg01} Norberg, P., et al.\ 
2001, MNRAS, 328, 64
\bibitem[Norberg et al.(2002)]{N02} Norberg, P., et al.\ 
2002, MNRAS, 332, 827
\bibitem[Peacock \& Smith(2000)]{PS00} Peacock, J.~A., \& Smith, R.~E.\ 2000, MNRAS, 318, 1144
\bibitem[Peebles(1980)]{P80} Peebles, P.~J.~E.\ 1980, 
Research supported by the National Science Foundation.~Princeton, N.J., 
Princeton University Press, 1980.~435 p.
\bibitem[Ross et al.(2006)]{R06} Ross, A.~J., Brunner, 
R.~J., \& Myers, A.~D.\ 2006, ApJ, 649, 48
\bibitem[Ross et al.(2007)]{R07} Ross, A.~J., Brunner, 
R.~J., \& Myers, A.~D.\ 2007, ApJ, 665, 67
\bibitem[Schlegel, Finkbeiner \& Davis(1998)]{Sc} Schlegel, D.~J., 
Finkbeiner, D.~P., \& Davis, M.\ 1998, ApJ, 500, 525
\bibitem[Scoccimarro et al.(2001)]{scoc01} Scoccimarro R., Sheth R. K., Hui L., Jain B., 2001, ApJ, 546, 20
\bibitem[Scoville et al. (2006)]{Scoville06} Scoville, N., et al.\ 2007, ApJS, 172, 150
\bibitem[Scranton et al.(2002)]{Scr02} Scranton, R., et al.
2002, ApJ, 579, 48 
\bibitem[Seljak et al.(2005)]{Sel05} Seljak, U., et al.\ 
2005, Phys. Rev. D, 71, 043511
\bibitem[Sheth \& Tormen(1999)]{Sheth99} Sheth, R.~K., \& 
Tormen, G.\ 1999, MNRAS, 308, 119
\bibitem[Sheth et al.(2001)]{Sheth01} Sheth, R.~K., Mo, H.~J., 
\& Tormen, G.\ 2001, MNRAS, 323, 1
\bibitem[Skibba et al.(2008)]{Skibba08} Skibba, R.~A., et al.\ 
2008, arXiv:0811.3970  
\bibitem[Smith et al.(2003)]{Smith} Smith, R.~E., et al.\ 
2003, MNRAS, 341, 1311  
\bibitem[Tinker et al.(2005)]{Tinker} Tinker J. L., Weinberg D. H., Zheng Z., Zehavi I., 2005, ApJ, 631, 41
\bibitem[Tinker et al.(2008)]{Tink08} Tinker, J.~L., Conroy, 
C., Norberg, P., Patiri, S.~G., Weinberg, D.~H., 
\& Warren, M.~S.\ 2008, ApJ, 686, 53
\bibitem[Wake et al.(2008)]{Wake08} Wake D.~A. et al., \ 2008 MNRAS, 387, 1045
\bibitem[Willmer et al.(1998)]{W98} Willmer, C.~N.~A., da 
Costa, L.~N., \& Pellegrini, P.~S.\ 1998, AJ, 115, 869
\bibitem[York et al.(2000)]{Y} York, D.~G., et al. 2000, 
AJ, 120, 1579 
\bibitem[Yee et al.(2005)]{Yee05} Yee, H.~K.~C., Hsieh, 
B.~C., Lin, H., \& Gladders, M.~D.\ 2005, ApJl, 629, L77 
\bibitem[Zehavi et al.(2004)]{Z04} Zehavi, I., et al.\ 
2004, ApJ, 608, 16
\bibitem[Zehavi et al.(2005)]{Z05} Zehavi, I., et al.\ 
2005, ApJ, 630, 1 
\bibitem[Zheng et al.(2005)]{zheng05} Zheng, Z., et al.\ 2005, ApJ, 633, 791
\bibitem[Zheng et al.(2007)]{ZZ07} Zheng, Z., Coil, A.~L., 
\& Zehavi, I.\ 2007, ApJ, 667, 760


\end{thebibliography}
\end{document}